\documentclass[conference,a4paper]{IEEEtran}
\usepackage{xcolor}
\usepackage{balance}

\ifCLASSINFOpdf
  \usepackage[pdftex]{graphicx}
\else
  \usepackage[dvips]{graphicx}
\fi

\ifCLASSOPTIONcompsoc
 \usepackage[caption=false,font=normalsize,labelfont=sf,textfont=sf]{subfig}
\else
 \usepackage[caption=false,font=footnotesize]{subfig}
\fi

\begin{document}

\title{State of the Art on Stacked Intelligent Metasurfaces \\ {\huge{Communication, Sensing and Computing in the Wave Domain}}}

\author{\IEEEauthorblockN{
Marco Di Renzo
}                                     
\IEEEauthorblockA{Universit\'e Paris-Saclay, CNRS, CentraleSup\'elec \\ Laboratoire des Signaux et Syst\'emes, 91192 Gif-sur-Yvette, France}
\IEEEauthorblockA{ \emph{marco.di-renzo@universite-paris-saclay.fr} }
}

\maketitle

\begin{abstract}
Stacked intelligent metasurface (SIM) is an emerging technology that capitalizes on reconfigurable metasurfaces for several applications in wireless communications. SIM is considered an enabler for integrating communication, sensing and computing in a unique platform. In this paper, we offer a survey on the state of the art of SIM for wireless communications.
\end{abstract}

\begin{IEEEkeywords}
Dynamic metasurfaces, reconfigurable intelligent surfaces, stacked intelligent metasurfaces.
\end{IEEEkeywords}

\section{Introduction}
In the last few years, dynamic (reconfigurable) metasurfaces, often referred to as reconfigurable intelligent surfaces (RISs) in wireless communications, have received major attention for several telecommunication applications \cite{9140329}, \cite{9326394}, \cite{9864116}. RISs can be utilized to customize the propagation of radio waves through smart reflections and refractions \cite{10555049}, \cite{DBLP:journals/corr/abs-2403-16458}, as well as to realize low-complexity transceivers, e.g., to modulate data for backscattering communications \cite{9749219}, \cite{s11432-022-3626-5}, \cite{articleScienceChina}.

More recently, reconfigurable metasurfaces have been proposed to conceive a new technology in which multiple dynamic metasurfaces are stacked together, in order to process data by mimicking a feed-forward artificial neural network that operates in the electromagnetic domain. In the context of wireless communications, this technology is referred to as stacked intelligent metasurface (SIM). The main advantage of SIM is to reduce the power consumption and complexity of digital artificial neural networks, as it is known that the power consumption of analog-to-digital converters increases exponentially with the number of bits and linearly with the bandwidth \cite{7258468}. Therefore, a wave-domain implementation is often considered an energy efficient alternative to fully-digital solutions \cite{10515204}. It is worth mentioning that an SIM can be viewed as an instance of a so-called reconfigurable deep diffractive neural network \cite{doi:10.1126/science.aat8084}, \cite{articleNatureE}, \cite{10.1063/5.0191977}, which capitalizes on the properties of nearly-passive reconfigurable metasurfaces for operation at different frequency bands \cite{articleMonticone} (not necessarily in the optical domain). Specifically, SIM is considered a multi-function technology for integrating communication, sensing and computing and for implementing them efficiently in the wave domain.

\section{State of the Art}
In this section, we classify the research works available on SIM with focus on applications for wireless communications.

\subsection{Beamforming Design}
Most available research works are focused on beamforming design, with the objective of optimizing and comparing the performance of SIM against fully-digital implementations. Examples of recent contributions include \cite{10515204}, \cite{10158690}, \cite{10279173}, \cite{10379500}, \cite{10534211}, \cite{10683447}, \cite{DBLP:journals/corr/abs-2403-18307}, \cite{10679332}, \cite{darsena2024designstackedintelligentmetasurfaces}, \cite{10445200}, \cite{DBLP:journals/corr/abs-2402-16405}, \cite{rezvani2024uplinkwavedomaincombinerstacked}. In \cite{10379500}, for example, several illustrations about the impact of the number of stacked layers are presented. Machine learning methods have been utilized as well \cite{DBLP:journals/corr/abs-2408-04837}, \cite{10622385} \cite{yang2024jointsimconfigurationpower}. Also, the synergy between SIM and a cell-free network architecture has been analyzed in \cite{10535263}, \cite{10679324}. The optimization of SIM considering the energy efficiency as optimization criterion has been recently tackled in \cite{DBLP:journals/corr/abs-2409-00628}. The application of SIM for physical layer security has been recently studied in \cite{10679315}.

\subsection{Channel Estimation}
Channel estimation has been the subject of intense research for RIS-aided communications \cite{9847080}. A few research works are available for SIM-aided communications as well \cite{10445164}, \cite{10679297}, \cite{10571026}. It needs to be mentioned, however, that channel estimation is SIM-aided systems is similar to typical RIS-aided networks only if the reconfigurable surfaces are located throughout the environment and are not an integral part of the transceivers. If an SIM is part of the transmitter, then the channel estimation process is easier, and is more similar to conventional  multiple-antenna systems. The number of parameters to estimate is, however, higher.

\subsection{Direction of Arrival Estimation}
One of the key features of SIM compared to other technologies is the ability to perform complex computations in the wave domain. A notable example is to process the incident signals for estimating their direction of arrival \cite{10557708}, \cite{10622963}. In \cite{10557708}, for example, the authors have proved that an SIM can be configured to compute the bi-dimensional discrete Fourier transform of an input signal, which can be utilized to estimate the direction of arrival.

\subsection{Integrated Sensing and Communications}
Motivated by the potential of SIM for implementing beamforming and direction of arrival estimation in the wave domain, a few recent works have evaluated the possibility of jointly combining these two functions \cite{DBLP:journals/corr/abs-2407-03566}, \cite{10643881}, \cite{DBLP:journals/corr/abs-2405-01104}, \cite{li2024transmitbeamformingdesignisac}. For example, the authors of \cite{DBLP:journals/corr/abs-2405-01104} have shown that increasing the number of layers of an SIM is beneficial for jointly improving communication and sensing performance.

\subsection{Semantic Communications}
Due to the powerful information processing capabilities in the wave domain, SIM has recently been proposed for implementing semantic encoders and decoders that are jointly designed to account for the content to be transmitted at a lower complexity and power consumption compared to full-digital solutions \cite{DBLP:journals/corr/abs-2407-15053}.

\subsection{Electromagnetic Modeling}
Most research works currently available in the literature utilize a simple communication model: each layer of the SIM is characterized by unit-amplitude transmission coefficients and the inter-layer propagation channel is modeled based on scalar Rayleigh-Sommerfield's diffraction theory. Recently, some authors have proposed multiport network theory \cite{10574199}, as a more accurate approach for modeling the near-field channel among all the layers of the SIM \cite{10530995}.

\subsection{Experimental Evaluation}
Even though some research works in the context of deep diffractive neural networks have validated the feasibility of wave-domain computing based on the cascade of multiple diffractive layers, these research works are mainly focused on optical systems \cite{doi:10.1126/science.aat8084}, \cite{10.1063/5.0191977}, and in some cases they rely on active implementations, i.e., they require power amplifiers \cite{articleNatureE}. In the context of wireless communications, the authors of \cite{DBLP:journals/corr/abs-2405-01104} have recently reported some experimental results on an SIM implementation based on one-bit dynamic metasurfaces for both communication and sensing applications. The preliminary results reported in \cite{DBLP:journals/corr/abs-2405-01104} highlight the importance of carefully optimizing an SIM in order to take advantage of the multi-layer architecture.

\balance

\section{Conclusion}
In this paper, we have provided a short summary of the state of the art on SIM. The survey has highlighted that several research works have been published in a relatively short time, and that several open research issues remain to be investigated. For example, the vast majority of research works are focused on beamforming optimization, while fewer works have analyzed the powerful wave-domain processing capabilities of SIM for information processing. Electromagnetic models as well as full-wave simulations and hardware implementations deserve more attention from the research community. A video presentation on SIM is available in \cite{VIDEO}.

\section*{Acknowledgment}
This work was supported in part by the European Commission through the Horizon Europe project COVER under grant agreement 101086228, the Horizon Europe project UNITE under grant agreement 101129618, and the Horizon Europe project INSTINCT under grant agreement 101139161, as well as by the Agence Nationale de la Recherche (ANR) through the France 2030 project ANR-PEPR Networks of the Future under grant agreements NF-PERSEUS 22-PEFT-0004, NF-YACARI 22-PEFT-0005, NF-SYSTERA 22-PEFT-0006, NF-FOUND 22-PEFT-0010, and by the CHIST-ERA project PASSIONATE under grant agreements CHIST-ERA-22-WAI-04 and ANR-23-CHR4-0003-01.

\bibliographystyle{IEEEtran}
\bibliography{biblio}

\end{document}